\documentclass[galaxies,article,accept,moreauthors,pdftex]{Definitions/mdpi} 

\firstpage{1} 
\pubvolume{1}
\issuenum{1}
\articlenumber{0}
\pubyear{2021}
\copyrightyear{2020}
\datereceived{} 
\dateaccepted{} 
\datepublished{} 
\hreflink{https://doi.org/} % If needed use \linebreak

\Title{Simulating the enrichment of fossil radio electrons by multiple radio galaxies}
\TitleCitation{Fossil electrons by radio galaxies}

\newcommand{\enzo}{\it{\small ENZO}}
\newcommand{\magcow}{\it{\small MAGCOW}}
\Author{Franco Vazza $^{1,2,3}$,Denis Wittor$^{3,1}$, Marcus Br\"{u}ggen$^{3}$, Gianfranco Brunetti$^{2}$}
\AuthorNames{Franco Vazza et al.  }
\AuthorCitation{Vazza F. et al.}
\address{%
$^{1}$ \quad Dipartimento di Fisica e Astronomia, Universit\'{a} di Bologna, Via Gobetti 93/2, 40129 Bologna, Italy\\
$^{2}$ \quad INAF-Istitituto di Radio Astronomia, via Gobetti 101, 40129 Bologna, Italy\\
$^{3}$ \quad Hamburger Sternwarte, Universit\"{a}t Hamburg, Gojenbergsweg 112, 41029 Hamburg, Germany\\}
\corres{Correspondence: franco.vazza2@unibo.it}

\abstract{We simulate the evolution of relativistic electrons injected into the intracluster medium by five radio galaxies. We study  the spatial transport and the emission properties of the injected radio plasma over a $\sim 5$ Gyr period, and the sequence of cooling and re-acceleration events experienced by electrons, using a Lagrangian approach joint with a numerical method to model the evolution of momentum spectra of relativistic electrons. When compared with electrons injected by shock waves, electrons injected by radio galaxies (here limited to a single injection event) in our tests are unable to fuel large, $\sim \rm ~Mpc$ sized radio relics with fossil electrons, as required by current theoretical models, while electrons previously seeded by other shocks can do this.  On the other hand, the combination of seeding from radio galaxies, and of re-acceleration events from plasma perturbation, can produce detectable, small scale and filamentary emissions in the proximity ($\leq 100-200$ kpc) of radio galaxies.} 

\begin{document}
\section{Introduction}
Radio galaxies are a spectacular example of how accretion discs surrounding supermassive black holes (SMBH) couple radically different astrophysical temporal and spatial scales: i.e. from the innermost accretion orbits of SMBH ($\leq 10^{-3} \rm ~pc$) to the expanding jets, recently observed out to $\sim 2 \rm ~Mpc$ from their host galaxy \citep[e.g.][]{2022A&A...660A...2O}.
In galaxy clusters and groups, radio galaxies represent a prominent and visible reservoir of magnetic fields and non-thermal particles \cite[e.g.][]{Volk&Atoyan..ApJ.2000,hardcastlecroston}. 
The recent decade or so suggested that a volume-filling distribution of fossil relativistic electrons is often required in order to explain the observed radio power of radio relics, i.e. elongated and polarised radio sources that are typically associated with merger shock waves \cite[e.g.][for a review]{2019SSRv..215...16V}. Their low Mach numbers render direct Fermi I acceleration from the thermal pool challenging  \cite[e.g.][]{ka12, 2013MNRAS.435.1061P,2020A&A...634A..64B,2021ApJ...914...73Z}. The presence of a volume filling distribution of "fossil" ($\gamma \leq 10^3$, where $\gamma$ is the Lorentz factor) has been suggested to explain the formation of radio halos via Fermi II turbulent re-acceleration  \cite[e.g.][for a recent review]{bj14}. Even more recently it has been invoked to explain the large-scale emission detected at the extreme periphery of a few interacting clusters of galaxies \cite[e.g.][]{2019Sci...364..981G,bv20} or on much larger scales than the classic radio halos \cite[e.g.][]{2022Natur.609..911C,2022SciA....8.7623B}.

Tails of radio-emitting plasma from radio galaxies mix with diffuse radio emission from the ICM \cite[e.g.][]{2017PhPl...24d1402J,nolting19a}. Recent low-frequency observations are detecting complex morphologies of remnant plasma, injected by radio galaxies and in different stages of mixing their non-thermal content with their surrounding environment\cite[e.g.][]{2017SciA....3E1634D,2018MNRAS.473.3536W,2020A&A...634A...4M,2022MNRAS.514.3466Q,brienza22}.

The new generation of radio telescopes (e.g. LOFAR, MWA, ASKAP, MEERKAT) is capable of producing an unprecedente view of  this phenomenon. At the same time, there presently are only a few simulations designed to connect diffuse re-acceleration mechanisms in the ICM to the ageing of radio lobes. %Hence, it will not be possible to accurately infer the age of such sources, and to use radio observations to time the epoch and duty cycle of AGN feedback events that fill the reservoir of cosmic-ray electrons in the ICM. 

Simulating the multi-scale and long term interplay between radio jets close and cluster weather still is a challenge, even for modern numerical simulations, which typically have to focus on specific ranges of spatial/temporal scales and physical aspects of the problem. This is particularly demanding for cosmological simulations, in which properly resolving the accretion regions of supermassive black holes is impossible. 
To give a few examples, 
\citet[][]{xu09} and \cite{2011ApJ...739...77X} included magnetised outflows from radio galaxies in cosmological {\enzo} simulations, and studied the build-up of cluster magnetic fields from the injection of individual AGN jets, but there was no physical coupling between accretion and feedback powers. 
\citet{2012ApJ...750..166M} used an Eulerian Magneto Hydrodynamical approach to follow in detail the effect of "cluster weather" on radio lobes, by injecting magnetised jets into a galaxy cluster extracted from a different (Smoothed Particle Hydrodynamics, SPH) cosmological simulation.  \citet{2020arXiv200812784B} used the moving mesh AREPO code to realise high-resolution simulations of jets and predict X-ray and radio properties, but had to assume in post-processing distributions of magnetic fields and relativistic electrons. 

More recently, the complex semi-analytic modelling of the radio emission from jets leading to Fanaroff Riley I and II systems have been presented by \citet{turner23}, and this model has been incorporated in simulations of jets released in the atmospheres of clusters of galaxies, extracted from a large set of SPH cosmological simulations by \citet{2022arXiv221210059Y}. As for similar previous work \citep[e.g.][]{2012ApJ...750..166M} the usage of a composite resimulation allows to greatly increase the physical realism that can be modelled in the interaction of jets with their environment; however the same strategy prevents to follow the evolution of lobe materials for timescales longer than $\sim 100-200$ Myr since the injection, as it neglects the long term effect of gas self-gravity, as well further cluster perturbation from outside the resimulation region.

%...
Our series of recent works on this topic  \citep[][]{va21b,va22} aims at producing a complementary view on the above studies, by implementing the macroscopic effects of radio jets at a coarser spatial resolution, but also following the spatial and spectral energy evolution of the relativistic particles over billions of years after their injection, in fully cosmological MHD simulations. 
 First applications of our simulations to real radio sources have recently been presented in \cite{Hodgson21a}, \cite{2021Galax...9...93V} and \cite{brienza22}.

In this paper, we apply the same framework to study the spatial transport of electrons injected by five radio sources, all with different feedback powers (in our case, initially assigned as thermal and magnetic power). These radio galaxies are activated at the same time, within an already formed $\sim 10^{14} M_{\odot}$ cluster of galaxies. 
With this new analysis, we mainly want to address the following questions:
\begin{itemize}
    \item Under which conditions can radio galaxies fill the ICM  with relativistic electrons to a significant degree?
    \item Is the predicted distribution of re-accelerated fossil relativistic electrons suitable to produce diffuse radio emission, and is this compatible with radio observations? 
\end{itemize}

Our paper is structured as follows:
In Section~2, we present the cosmological simulations and all numerical methods employed in this paper. In Section~3, we give our main results, first focusing on the evolution of electrons seeded by radio jets (Section~3.1), and then by comparing with the evolution of electrons injected by merger shocks (Section~3.2). Our conclusions are given in Section~4. 

Throughout this paper, we use the following cosmological parameters: $h = 0.678$, $\Omega_{\Lambda} = 0.692$, $\Omega_{\mathrm{M}} = 0.308$ and $\Omega_{\mathrm{b}} = 0.0478$, based on the results from \cite{2016A&A...594A..13P}.

\section{Methods \& Material}
\label{methods}

%All methods and algorithms used to produce this new simulation have been recently presented in detail in Pap I and Pap II. 

In order to make this section as compact as possible, we highlight only the key technical aspects of our simulations, while referring the reader to our previous publications \cite{va21b,va22} for the technical details. 

\subsection{Cosmological Simulations \& Feedback from Active Galactic Nuclei}

We produced cosmological, adaptive mesh refinement {\enzo}-MHD \cite{enzo14} simulations using the MHD solver with a Lax-Friedrichs  Riemann solver to compute the fluxes in the Piece-wise Linear Method, combined with the Dedner cleaning method \cite[][]{ded02} implemented by \cite{wa09}. We start from a simple uniform magnetic field at $z=50$ with a value of $B_0=0.1  \ \mathrm{nG}$ in each direction. This simplistic model has been shown to produce radio signature which are in tension with a few radio observations of the low density cosmic web, making tangled initial magnetic fields preferred \citep[][]{va21magcow}, yet the differences expected within clusters of galaxies are negligible, owing to the effect of small-scale dynamo amplification \citep[][]{do99,review_dynamo,va21}.

The total simulated volume is of (50 Mpc$/h$)$^3$ and is sampled with a root grid of $128^3$ cells and dark matter particles. We further use four additional nested regions with increasing spatial resolution (up to $\Delta x=24.4$~kpc$/h$)  within a cubic (5.2 Mpc)$^3$ sub-volume, where a $M_{100} \approx 1.5 \cdot 10^{14} M_\odot$ cluster forms. 
During run-time, two additional levels of mesh refinement are added using a local gas/DM overdensity criterion ($\Delta \rho/\rho \geq 3$), allowing the simulation to reach a maximum resolution of $\approx 6 \rm ~kpc/h= 8.86$~kpc.  As a result of our nested grid approach, the mass resolution for dark matter in our cluster formation region is of  $m_{\rm DM}=2.82 \cdot 10^{6} ~M_{\odot}$ per dark matter particle, for the highest resolution particles that are used to fill the innermost AMR level since the start of the simulation. No radiative gas cooling or other galaxy formation-related physics (e.g. star formation or feedback from supernovae) was included, beside the single jet feedback episode from eight different radio galaxies, explained in the next Subsection. 

The thermal and magnetic properties of this cluster have been described in detail in \citet{va21b} and \citet{va22}, and are largely independent of the specific kind of AGN feedback. 
The galaxy cluster has a virial mass $M_{100} \approx 1.1 \cdot 10^{14} M_\odot$ at $z=0.5$, which grows to $M_{100} \approx 1.5 \cdot 10^{14} M_\odot$ by $z=0$ through the accretion of subsequent minor mergers, the most prominent ones being at  $z \sim 0.3$ and $z \sim 0.1$ and mostly directed along one of the diagonals of the computational box. 

At $z=0.5$, we identify the five most massive halos in the innermost high resolution $5.2^3 \rm ~Mpc^3$ volume, in the $\sim  10^{9}-10^{11} M_\odot$ range,   and place a SMBH particle at the center of each of them. We assume a black hole mass of $M_{\rm BH}=0.01 ~M_{\rm 200}$, where $M_{\rm 200}$ is the total (gas+DM) mass of each halo \footnote{Unlike in most cosmological simulations, we do not start from seed SMBH particles with a very low mass at high redshift, and monitor its growth in time \cite[e.g.][]{2011ApJ...738...54K}, but instead we place them a late redshift, and assign them already a mass compatible with the prediction of the range of masses of SMBH in fully formed and massive galaxies.}

\begin{table}
\begin{center}
\caption{Main parameters of SMBH and jet/radio power associated with the jets they emit (in which case the power is the sum of the power in the two jets).}  
\footnotesize
\centering \tabcolsep 2pt
\begin{tabular}{c|c|c|c|c|c}
 ID & $M_{\rm BH}$  & $\dot{M}_{\rm BH}/\dot{M}_{\rm Edd}$ & $P_j$ & $L_{140}$ \\  
  & [$M_{\odot}$] & & [erg/s] & [erg/s/Hz ]\\ \hline  
A & $8.7 \cdot 10^{8}$   & 0.0026 & $1.4 \cdot 10^{43}$  & $3.4 \cdot 10^{32}$ &  \\
B & $3.5 \cdot 10^{7}$ & 0.0004 & $1.0 \cdot 10^{41}$ & $1.6 \cdot 10^{31}$ \\
C & $1.4 \cdot 10^{7}$ & 0.00004 & $1.1 \cdot 10^{40}$ &  $5.3 \cdot 10^{30}$\\
D & $1.3 \cdot 10^{7}$ & 0.00007 & $1.3 \cdot 10^{39}$ &$2.9 \cdot 10^{31}$\\
E & $8.1 \cdot 10^{6}$ & 0.00004 & $7.1 \cdot 10^{38}$ &$1.9 \cdot 10^{31}$\\
  \end{tabular}
  \end{center}
\label{table:tab1}
\end{table}

\begin{figure*}
\begin{center}
\includegraphics[width=0.5\textwidth]{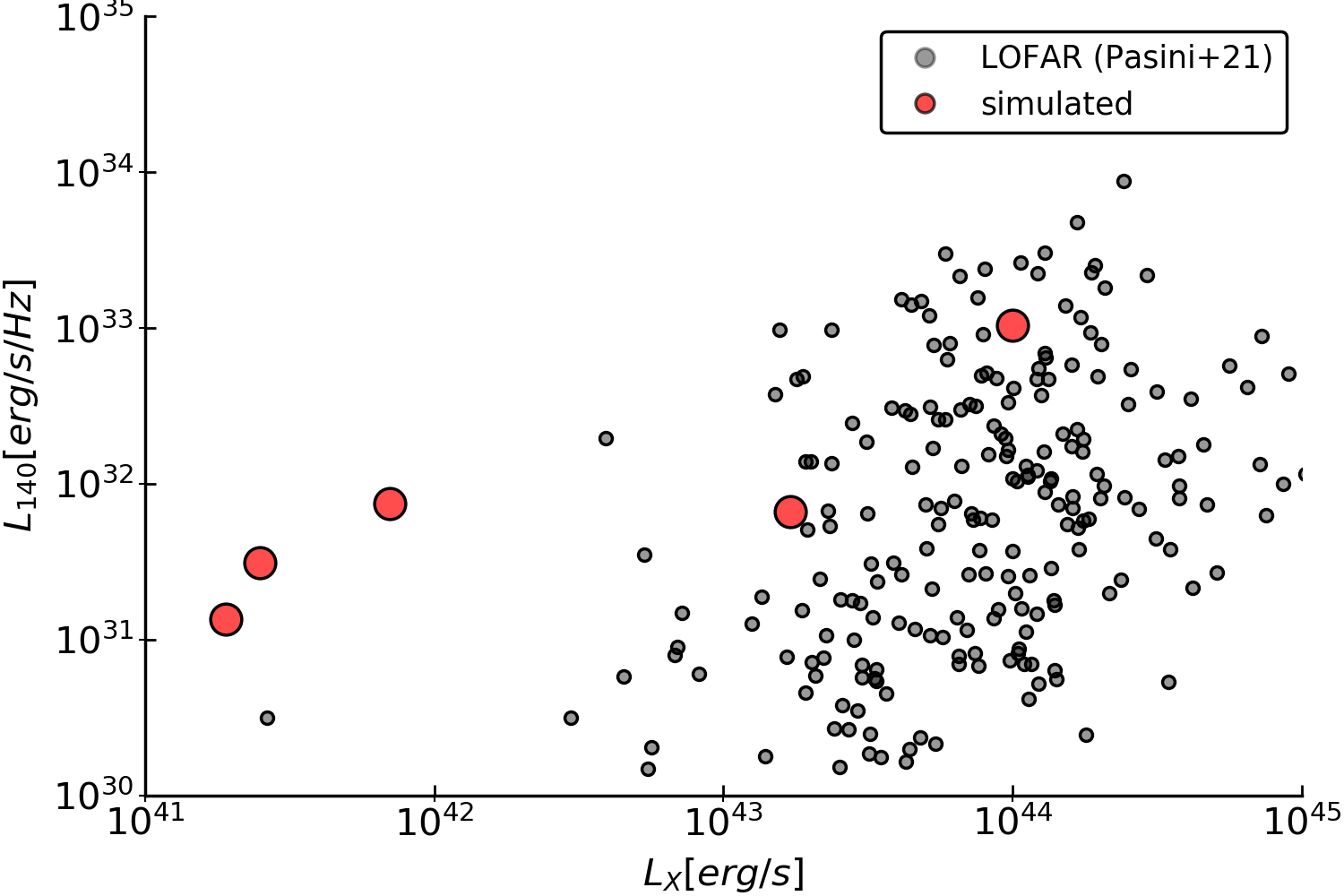}
\end{center}
\caption{Total radio emission within $\leq 500 \rm ~kpc$ from each simulated radio galaxy at 
 140MHz, versus the X-ray emission in the [0.5-2]keV range within the same regions in the host cluster. The grey points show the real LOFAR-HBA and eRosita observations from \cite{pasini21}.}.
\label{fig:pasini}
\end{figure*}  

SMBH particles have initial masses in the $8.1 \cdot 10^{6}-8.7 \cdot 10^{8} M_{\odot}$ range, and after injection we estimate their gas accretion rate,  based on the usual Bondi–Hoyle formalism:

\begin{equation}
    \dot{M}_{\rm BH}=\frac{4 \pi \alpha_B G^2 M_{\rm BH}^2\rho}{(v_{\rm rel}^2+c_s^2)^{3/2}} ,
    \label{eq:bondi}
\end{equation}
 in which $\rho$ is the local gas density,  $v_{\rm rel}$ is the SMBH velocity relative to the gas, $c_s$ is the sound speed at Bondi radius (which we assume to be relative to a fixed $10^6 ~\rm K$ temperature, considering that our simulation lacks the resolution and physical ingredients to predict the temperature in this region, also considering the lack of radiative cooling), and $\rho$ is the local gas density. $\alpha_B$ is a free normalisation factor, often employed in simulations incapable to resolve the Bondi radius \cite[e.g.][]{2009MNRAS.398...53B,gaspari12,2016Natur.534..218T},  which we here simply leave to $\alpha_B=1.0$ for all simulated SMBHs. 
 
 As shown in Tab. 1, our simulated SMBH have a range of accretion powers from $\sim 0.26\%$ to $\sim 0.0004\%$ of the Eddington limited accretion rate. In this sense they all represent a relatively quiet population of SMBH in the epoch when the jets are launched in the ICM. The quietness of our population can be partially ascribed to the absence of large mass accretion rates, which in turn follow from  the absence of radiative cooling (which
 tends to increase density and reduce the sound speed/temperature in the Eq.\ref{eq:bondi}); however this ensures that our results can represent rather average sources in clusters of galaxies.

 %This is at variance with our recent work \cite[][]{va22}, in which we used values from $\alpha_B=1$ to $50$, and is motivated by the fact that we want to study rather low power (and probably more frequent) radio jet events, and test if they are already enough to enrich the ICM with fossil electrons. 
 
The bolometric luminosity of each SMBH is the standard $L_{\rm BH} = \epsilon_r \dot{M}_{\rm BH} c^2$, where we assumed the standard values in the literature of $\epsilon_r=0.1$ for the radiative efficiency of the SMBH. This is used to compute the jet power released in the feedback stage (which only last, by construction, one root grid timestep, i.e. $\sim 32 \rm ~Myr$)  as  $P_j=\epsilon_r \epsilon_{\rm BH} \dot{M}_{\rm BH} c^2$, where $\epsilon_{\rm BH}$ is the factor that converts the bolometric luminosity to the thermal feedback energy (fixed to 0.05, also in line with what is done in most cosmological simulations).  

Jets are introduced at the same time ($z=0.495$) for all
SMBH particles, following  the same recipes presented in \cite{va22}. 
%Since the simulation neglects radiative cooling entirely, it cannot have a self-regulating mechanism to switch the feedback cycle on and off \cite[e.g.][for a few examples]{gaspari11b,2015ApJ...813L..17R,2019MNRAS.489..802R,2021MNRAS.504.3619T}, and therefore the 

 %This leads to accretion rates $\sim 3-6 \%$ of the Eddington limited accretion rates for our SMBH, yielding total jet powers of $P_j\sim 1-5 \cdot 10^{44} \rm erg/s$. 
 
SMBH particles inject $10\%$ of the total feedback energy as magnetic field energy (with the remaining 90\% injected as thermal energy, by introducing two magnetic loops ($2 \times 2$ cells at the highest resolution level), located at $\pm 1$ cell along the $z$-direction from each SMBH, i.e. at $\pm 8.86 \rm ~kpc$ separation). 
 Since the thermal feedback energy gets isotropically spread over a larger volume (i.e. 27 cells), the typical magnetic energy within the active jets is a factor of a few  larger than the kinetic and thermal energy within the same region, hence our jets are initially effectively magnetic energy dominated, at least in the first $\leq 100 \rm ~Myr$ since their creation. %In reality, we expect that also the cosmic-ray component should be important for the jet internal pressure of radio jets embedded in clusters \cite[e.g.][]{2018MNRAS.476.1614C}, yet in this simulation we can only track cosmic-ray electrons as a passive fluid, with no contribution to the dynamical pressure (see Section 6 for a discussion on the physical limitations of our model).

Even though imposing a fixed jet orientation along the $z$ axis for all our SMBH jets is artificial, in the long term the initial direction of jets is observed to play no role. We also tested in \citet{va22} that different choices in  the initial direction of jets produce no significant differences in the statistics of the thermodynamical properties of the ICM and of the related radio emission.

Our approach is similar to early work by \cite[][]{2006ApJ...643...92L} and \cite[][]{xu09}, who initialised toroidal magnetic fields on opposite sides of the SMBHs, showing that this procedure naturally leads to self-collimating outflows and a supersonic expansion in the cluster core. It shall be noticed that our chosen $2 \times 2$ cell configuration for the initial magnetic field loops is not enough to properly resolve the field loops in the model of \cite{2006ApJ...643...92L}; however this is not a source of concern because the focus of our work is more on the energetics and distribution of tracer particles, and not on the details of jet morphology.

 The distribution of the simulated radio power at 140 MHz, measured $\sim 32 \rm ~Myr$ since the start of the jet active stage and  computed after the injection of relativistic electrons and with the synchrotron approach explained in the next Section, is plotted in Fig. \ref{fig:pasini} as a function of the X-ray luminosity within $\leq 500 ~\rm kpc$ each source, to compare with the recent LOFAR-HBA observations by 
\citet{pasini21}. This suggests that the 
range of radio powers spanned by our sample is well within the one of observed real sources, even if our sample also includes non-central ones.  Recent work by \cite{2022MNRAS.513.3273S} used radio and X-ray observations of six rich clusters of galaxies to study the role of AGN feedback by non-central extended radio galaxies, and   concluded that while cluster-central sources are the dominant factor to balance cooling over the long-term, non-central sources also have a significant impact. 

\begin{figure*}
\begin{center}
\includegraphics[width=0.35\textwidth]{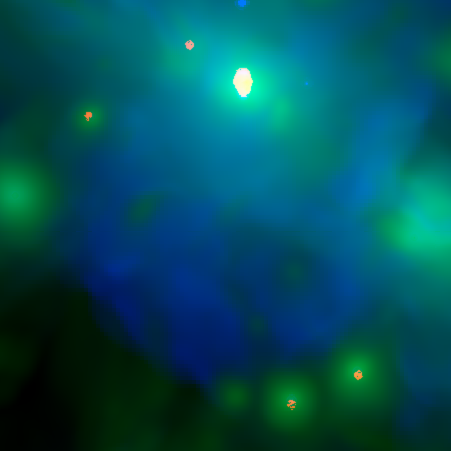}
\includegraphics[width=0.35\textwidth]{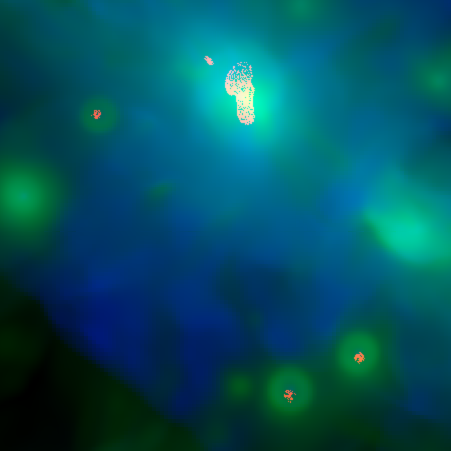}
\includegraphics[width=0.35\textwidth]{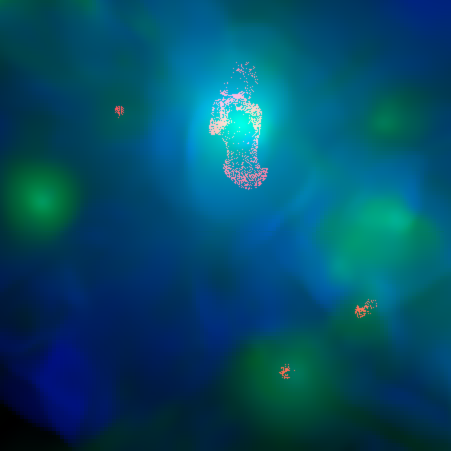}
\includegraphics[width=0.35\textwidth]{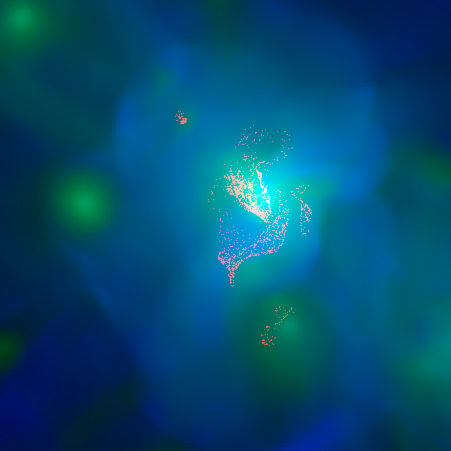}
\includegraphics[width=0.35\textwidth]{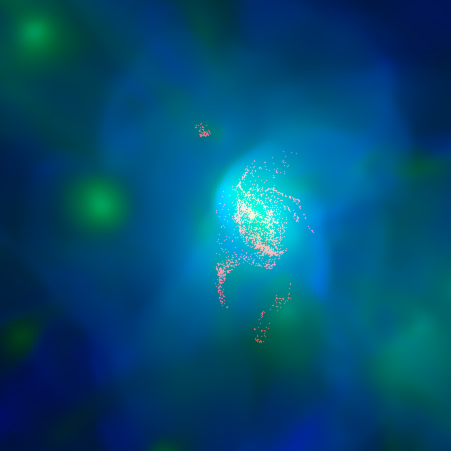}
\includegraphics[width=0.35\textwidth]{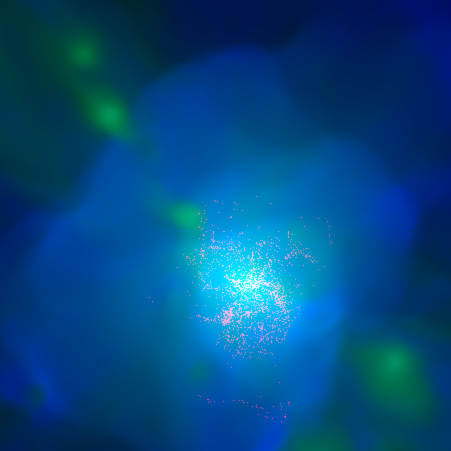}
\end{center}
\caption{Red-green-blue composite images showing the evolution of our simulated cluster of galaxies, and of the location of electron tracers injected by our simulated (5) radio galaxies. Each image is $5 \times 5$ Mpc$^2$ across. They show the mass-weighted gas temperature projected along the full line of sight (blue), the projected gas density (green) and the location of electron tracer, regardless of their energy (red). From the top left to the lower right panel, the redshifts are $z=0.493$, $0.452$, $0.382$, $0.201$, $0.102$ and $0.032$.}
\label{fig:movie1}
\end{figure*}

\subsection{Injection, tracking and energy evolution of relativistic electrons from radio galaxies} 
\label{subsec:brats}

We track the relativistic electrons with a post-processing Lagrangian scheme, running the CRATER code (\cite{wi16}) over all snapshots of our simulation ($221$ timesteps, with a time resolution of about $\Delta t \sim 32 \rm ~Myr$). 

We injected and propagated $\sim 5 \cdot 10^4$ particles, assigning them according to the local gas density distribution at $\leq 8 ~ \rm kpc$ from each active jets, after we discretized the tracer mass resolution to $m_{\rm trac}=8\cdot 10^5 ~M_{\odot}$. The various physical quantities of interest (gas density, velocity, divergence, vorticitiy, temperature and magnetic field intensity) are assigned to the tracers using a cloud-in-cell (CIC) interpolation method, while shocks were detected based on temporal temperature discontinuities. % Tracers also keep track of the local fluid divergence, $\nabla \cdot \vec{v}$, and of the fluid vorticity, $\nabla \times \vec{v}$, which serve as proxies for the local turbulence.

Electron spectra for each tracer were initialised such as to reproduce the observed radio spectra of radio galaxies, through the Continuous Injection-Off model \citep[e.g.][]{1994A&A...285...27K}, with initial normalisation following from assuming a $0.5\%$  fixed fraction between the injected relativistic electrons and the thermal proton density in the jet region \cite[e.g.][]{2012ApJ...750..166M}. For tests of this approach in our model we refer the reader to \citet{va22}. 

\bigskip
After initialisation, we solve for the time-dependent diffusion-loss equation of radio emitting particles with the parallel ROGER solver \footnote{https://github.com/FrancoVazza/JULIA/tree/master/ROGER}, using momentum bins equally spaced in $\rm log (p)$, in the $p_{\rm min} \leq p \leq p_{\rm max}$ momentum range (where $p=\gamma m_e v/(m_e c)$ is the normalised momentum of CR electrons). In all production runs, we used $p_{\rm min}=5$ and $p_{\rm max}=10^6$ and $d\log (p)=0.05$.

%Whenever present, injection from shocks (or initially from radio jets)  is treated as an instantaneous process owing to the time scales much shorter than the time step of our integration. 

We consider radiative losses, Coulomb losses and expansion (compression), as well as the acceleration from diffusive shock acceleration (DSA). The shock kinetic energy flux that we assumed to be converted into the acceleration of cosmic rays is: $\Psi_{\rm CR} = \xi_e ~\eta(\mathcal{M}) \rho_u (V_s^3 dx_t^2)/2$ ,
in which $\rho_u$ is the pre-shock gas density,  $V_s$ is the shock velocity and the combination $\xi_e ~\eta({\mathcal{M}})$ gives the CR electrons acceleration efficiency.

%For simplicity (and noting that most of shocks within the simulated ICM are quasi-perpendicular, and thus efficient, \cite[e.g.][]{wittor20,Banfi20}) we neglect the dependence of the shock acceleration efficiency on the local magnetic field topology, and only base the shock acceleration efficiency on the Mach number recorded by tracers. 

The latter is a combination of the energy going into cosmic rays, $\eta(\mathcal{M})$ (for which we use the convenient polynomial approximation by \cite{kj07}), and the electron-to-proton acceleration rate, $\xi_e$, which we fix by requiring an equal number density of supra-theral cosmic-ray electrons and protons above the injection momentum,  $\xi_e=(m_p/m_e)^{(1-\delta_{\rm inj})/2}$, consistent with \cite{2013MNRAS.435.1061P}. For simplicity, we neglect dependencies on the shock obliquity \citep[e.g.][]{2023MNRAS.519..548B}, on the basis that our previous work showed that the majority of internal shocks in the ICM are quasi-perpendicular and hence suitable for electron acceleration \cite[e.g.][]{wittor20,Banfi20}.

The injection momentum of electrons $P_{\rm inj}$ is linked to the thermal momentum of particles, i.e. $P_{\rm inj}= \xi P_{\rm th}$ ($P_{\rm th}=\sqrt{2 k_b T_d m_p}$), measured locally for each Lagrangian tracer. 

The working surface associated with each tracer, $dx_t^2$, is adjusted at run-time as $dx_t^3 = dx^3/n_{\rm tracers}$, where $dx^3$ is the volume initially associated with every tracer and ($n_{\rm tracer}$ is the number of tracers in every cell).

Supra-thermal electrons are injected by shocks with a power-law momentum distribution \citep[e.g.][]{sa99}, $Q_{\rm inj}(p) = K_{\rm inj} ~p^{-\delta_{\rm inj}} (1-p/p_{\rm cut})^{\delta_{\rm inj}-2}$ where the initial slope of the input momentum spectrum, $\delta_{\rm inj}$, follows from the standard DSA prediction, $\delta_{\rm inj} = 2 (\mathcal{M}^2+1)/(\mathcal{M}^2-1)$.  
$p_{\rm cut}$ is the cut-off momentum, which we set for every shocked tracer as the maximum momentum beyond which the radiative cooling time scale gets shorter than the acceleration time scale, $\tau_{\rm DSA}$: 
\begin{equation}
\tau_{\rm DSA} = \frac{3~D(E)}{V_s^2} \cdot \frac{r(r+1)}{r-1} ,
\label{eq:tDSA}
\end{equation}
 in which  $r$ is the shock compression factor and $D(E)$ is the electron diffusion coefficient as a function of energy, which  is poorly constrained as it depends on the microphysical conditions of the turbulent plasma. However, all plausible choices of $D(E)$ in Eq.~\ref{eq:tDSA} give an acceleration timescale which is many orders of magnitude smaller than the typical cooling time of radio emitting electrons \citep[e.g.][]{ka12}, hence we can asume the  momentum distribution at injection can be assumed to follow a power law within our momentum range of interest.

The normalisation factor for shock-injected electrons, $K_{\rm inj}$, follows from equating the cosmic ray energy flux crossing each tracer volume element, and the product between the total energy of cosmic rays ($E_{\rm CR}$) advected with a post-shock velocity ($v_d$): 
$\Psi_{\rm CR} ~dx_t = v_d E_{\rm CR}$ ($v_d$ is the post-shock velocity), similar to  \cite[][]{2013MNRAS.435.1061P}.

\begin{figure*}
\begin{center}
\includegraphics[width=0.95\textwidth]{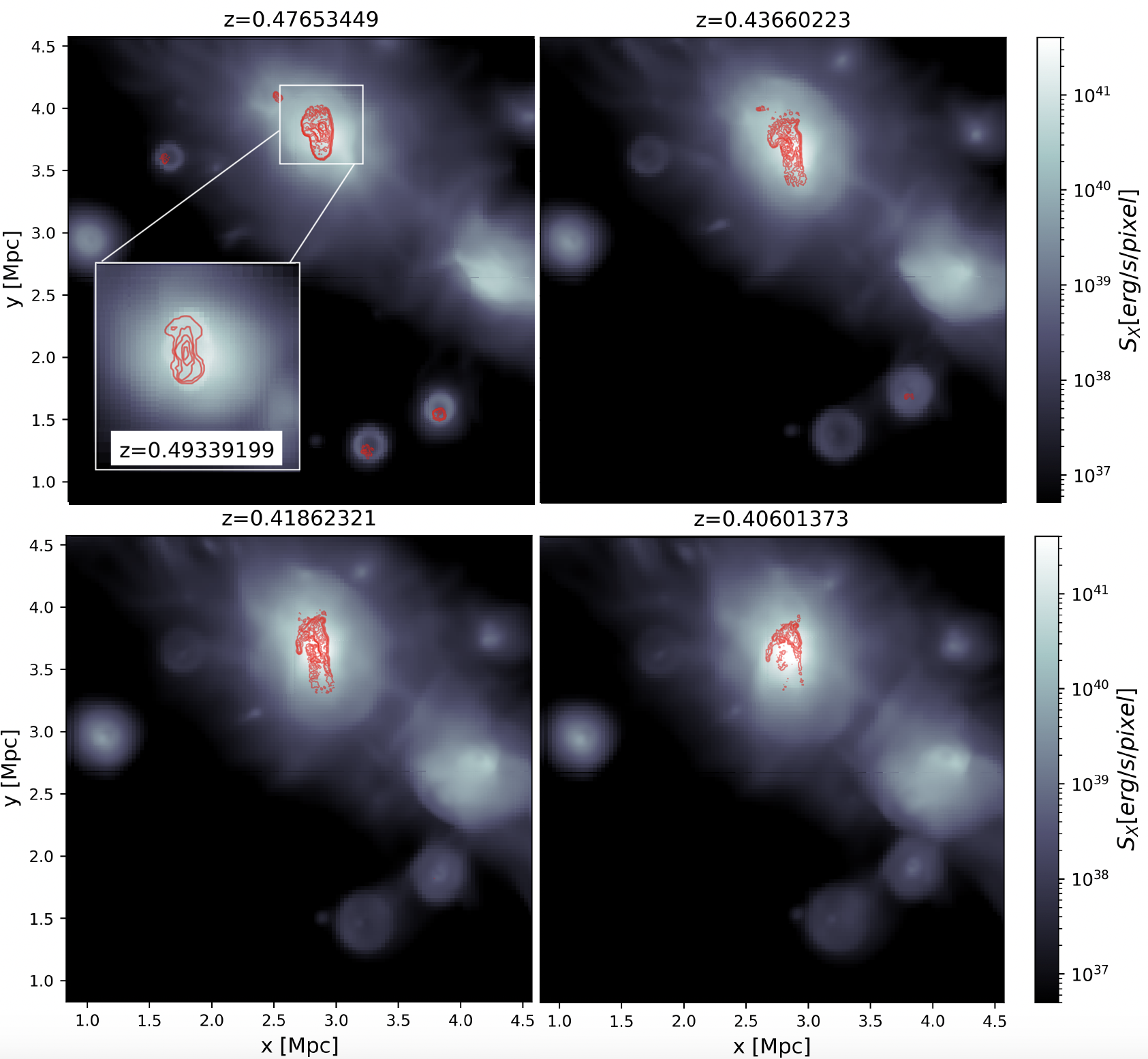}
%includegraphics[width=0.495\textwidth]{images/1mov1.png}
%\includegraphics[width=0.495\textwidth]{images/2mov1.png}
%\includegraphics[width=0.495\textwidth]{images/3mov1.png}
%\includegraphics[width=0.495\textwidth]{images/4mov1.png}
%\includegraphics[width=0.495\textwidth]{images/5mov1.png}
%\includegraphics[width=0.495\textwidth]{images/6mov1.png}

\end{center}
\caption{Evolution of the X-ray surface brightness (in the 0.5-2keV energy range, grey colors) and of the 
 radio emission at 50 MHz with LOFAR LBA, for all electrons seeded by radio jets at z=0.5 and later re-accelerated by shocks and turbulence (CST model, red contours, equally spaced by $\sqrt{2} \sigma$ and starting from $2 \sigma$, where $\sigma$ is the assumed LOFAR LBA sensitivity (see Sec. \ref{radio_jets}) for details). The inset in the first panel shows a zoom onto the most powerful radio galaxies at an earlier redshift, i.e. $\sim 64 \rm ~Myr$ since the injection of jets. The sequence continues in Fig.\ref{fig:movie2b} and the full movie can be watched at https://youtu.be/MPEFiIzr-V8}. 
\label{fig:movie2}
\end{figure*}  
%\begin{equation}
%E_{\rm CR} = \int_{p_{\rm inj}}^{p_{\rm cut}} Q_{\rm inj}(p) T(p) dp ,
%\end{equation}
%with $Q_{\rm inj}(p)$ defined as above and $T(p) = (\sqrt{1+p^2}-1)m_e c^2$. The integration yields 

%\begin{equation}
%E_{\rm CR} = \frac{K_{\rm inj} m_e c^2}{\delta_{\rm inj}-1} \left[\frac{B_x}{2} \left( \frac{\delta_{\rm inj}-2}{2},\frac{3-\delta_{\rm inj}}{2}\right) + p_{\rm cut}^{1-\delta_{\rm inj}} \left(\sqrt{1+p_{\rm cut}^2}-1  \right)  \right] , 
%\label{eq:ECR}
%\end{equation}
%where $B_x(a,b)$ is the incomplete Bessel function and $\rm x=1/(1+p_{\rm cut}^2)$, as in \cite[][]{2013MNRAS.435.1061P}. 

Additionally, we model re-acceleration by shocks  \cite[e.g.][]{2005ApJ...627..733M,kr11,ka12}: $ 
N(p)=(\delta_{\rm inj}+2) \cdot p^{-\delta_{\rm inj}} \int_{p_{min}}^p N_0(x) x^{\delta_{\rm inj}-1} dx$ 
where $\delta_{\rm inj}$ is the local slope within each energy bin and $N_0(p)$ is the input spectrum of electrons before the re-acceleration. 

We also include Fermi II re-acceleration via stochastic interaction with diffusing magnetic field lines in super-Alfvenic turbulence. The re-acceleration is computed following the Adiabatic Stochastic Acceleration (ASA) model \citep[][]{2016MNRAS.458.2584B}, and it depends on the amplitude of the local turbulent velocity, $\delta V_{\rm turb}$, measured from the gas vorticity, i.e. $\delta V_{\rm turb}= |\nabla \times \vec{v}| L$, for which we used the same fixed reference scale of $L \approx 27$~kpc to compute vorticity via finite differences (i.e. 3 cells on the high-resolution mesh).
The acceleration term for electrons is $dp/dt \approx p/\tau_{\rm ASA}$, for which 
$\tau_{\rm ASA}(p) = p^2/4~D_{\rm pp}$, which is in the $\tau_{\rm ASA} \sim 0.1-1 \rm ~Gyr$ range for the typical dynamics of these systems. 
 The diffusion coefficient in momentum space in this scenario is (see \cite{bv20} for the derivation) $D_{pp} \approx (48 F_{\rm turb} ~p^2)/(c~\rho_{ICM} ~V_A)$, which depends on the solenoidal turbulent energy flux (conserved in the Kolmogorov model of turbulence): $F_{\rm turb} = \frac{\rho \delta V^3_{\rm turb}}{2L}$.  For the sake of brevity, we omitted here the full derivation of the parameters above, in connection with the quantities recorded by our Lagrangian tracers, which can be found in longer detail in \cite{va22}. 

Finally, to account for the missing efficient  amplification of magnetic fields (due to the under-resolved role of small-scale dynamo) and avoid unphysical amount of turbulent reacceleration in the ASA scheme (since $\tau_{\rm ASA} \propto B$) we assign each tracer a  re-normalised estimate of the local magnetic field, in which we take the maximum between the magnetic field value directly produced by the MHD calculation, and the one assuming turbulent dynamo amplification, for a constant  $\eta_B=2\%$ conversion efficiency between the local turbulent kinetic energy flux of the solenoidal component and the energy density  of magnetic fields (see  \cite{va22} for detailed tests on the outcomes of this approach). 

The synchrotron emission from the electrons was computed as explained in \citet{va22}, i.e. using the (fast) fitting procedures introduced by \cite{2021MNRAS.507.5281C}, for five radio frequencies (50, 120, 610, 1400 and 5000 MHz). Our ROGER electron solver is designed to solve multiple populations of electrons under different physical scenarios. Hence we could simulate the outcome of different possible scenarios for the origin and the evolution of radio emitting electrons, in particular:

\begin{itemize}
\item {\it Electrons seeded by radio jets}: we followed the evolution of electrons injected by our radio galaxies at $z=0.5$, solely under the influence of loss processes (e.g. radiative processes, Coulomb collision, ionisation losses and adiabatic changes, model "C"), by including the additional injection of new electrons by diffusive shock acceleration, as well the re-acceleration by DSA on previously injected population of electrons (model "CS"), or by additionally including also the Fermi II re-acceleration by solenoidal turbulence (model "CST"). This set of models followed $\approx 3 \cdot 10^4$ tracers, all initially placed within the jet launching regions of our five radio galaxies. 

\item {\it Electrons seeded by merger shocks}: we initialised pools of tracers in the simulation at $z=0.4$ (i.e. well after the jet activity has ended for all considered radio sources), assigning them according to the gas density profile and giving them an initially negligible content of relativistic electrons. We let them evolve according to all loss processes, shock injection and shock/turbulent re-acceleration. This is similar to the previous "CST" physical model, with the important difference that the electrons are only seeded by merger/accretion shock waves after the short active stage of jets. This model is meant to check the effectiveness of the multiple-shock scenario explored in \citet{2022MNRAS.509.1160I}, in which a large fraction of the radio power from relics comes from shock re-acceleration of electrons previously injected by older structure formation shocks. In other words, this second model is meant to quantify the relative importance of fossil electrons injected by shocks, compared to fossil electrons injected by radio galaxies on longer timescales. This second set of models followed $\approx 2 \cdot 10^5$ tracers, placed in the cluster at $z=0.4$ following the cluster density profile and only limited to $T \geq 10^7 \rm K$ regions. 

\item {\it Prompt injection of electrons by shocks}: we  included a simplistic scenario in which we only compute the prompt injection of electrons, and their radio emission, at a single time-step and based on the DSA model outlined above, i.e. with injection efficiency scaling with the Mach number and normalisation depending on the shock kinetic energy flux. Otherwise the electron spectra are  deleted from one time step to the next. This last model is just meant to compare with the standard approach to predict radio emission from shocks in the simulated ICM, in which single snapshots of simulations are used, assuming quasi-stationary shock conditions and neglecting the presence of fossil electrons \citep[e.g.][]{sk08,ho08,va14relics}. This last scenario was run on top of the same distribution of tracers used for the previous model. 
\end{itemize}

\begin{figure*}
\begin{center}
    \includegraphics[width=0.49\textwidth,height=0.43\textwidth]{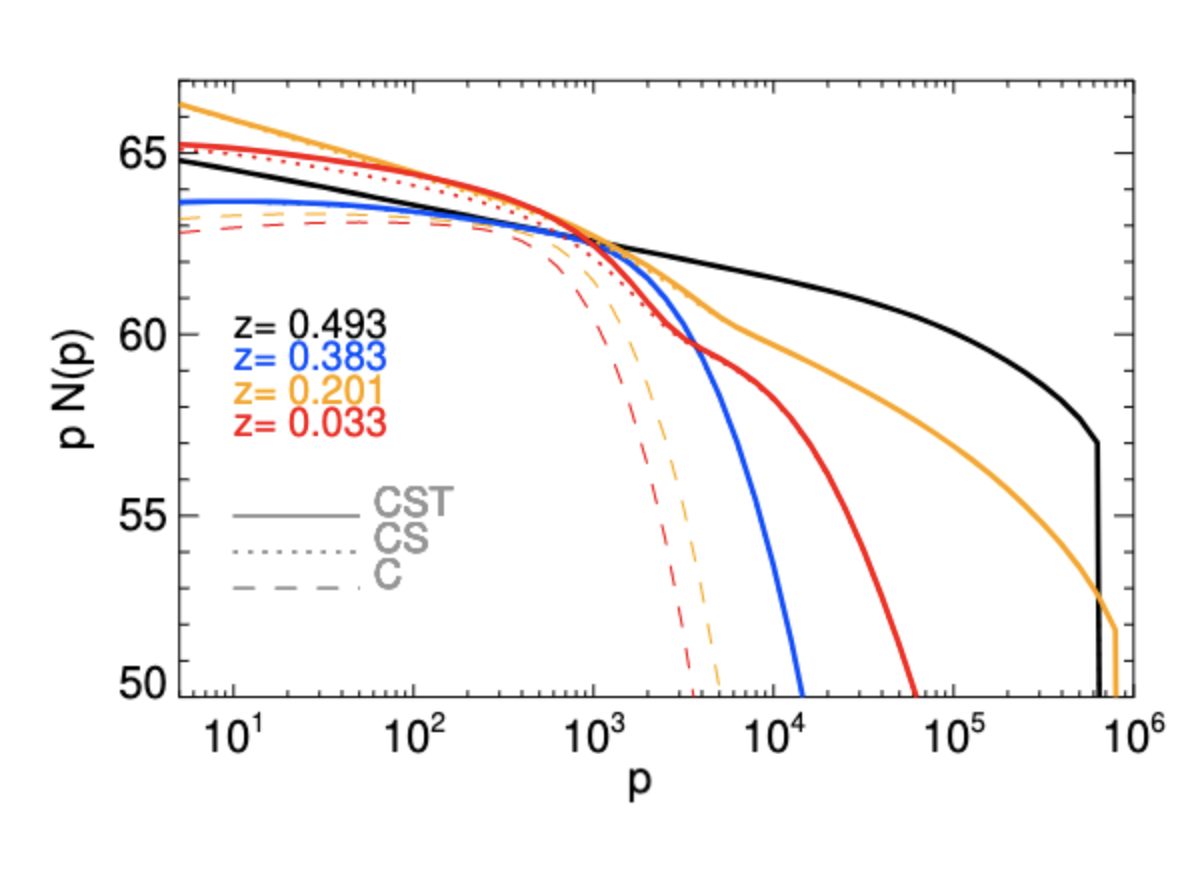}
     \includegraphics[width=0.45\textwidth,height=0.38\textwidth]{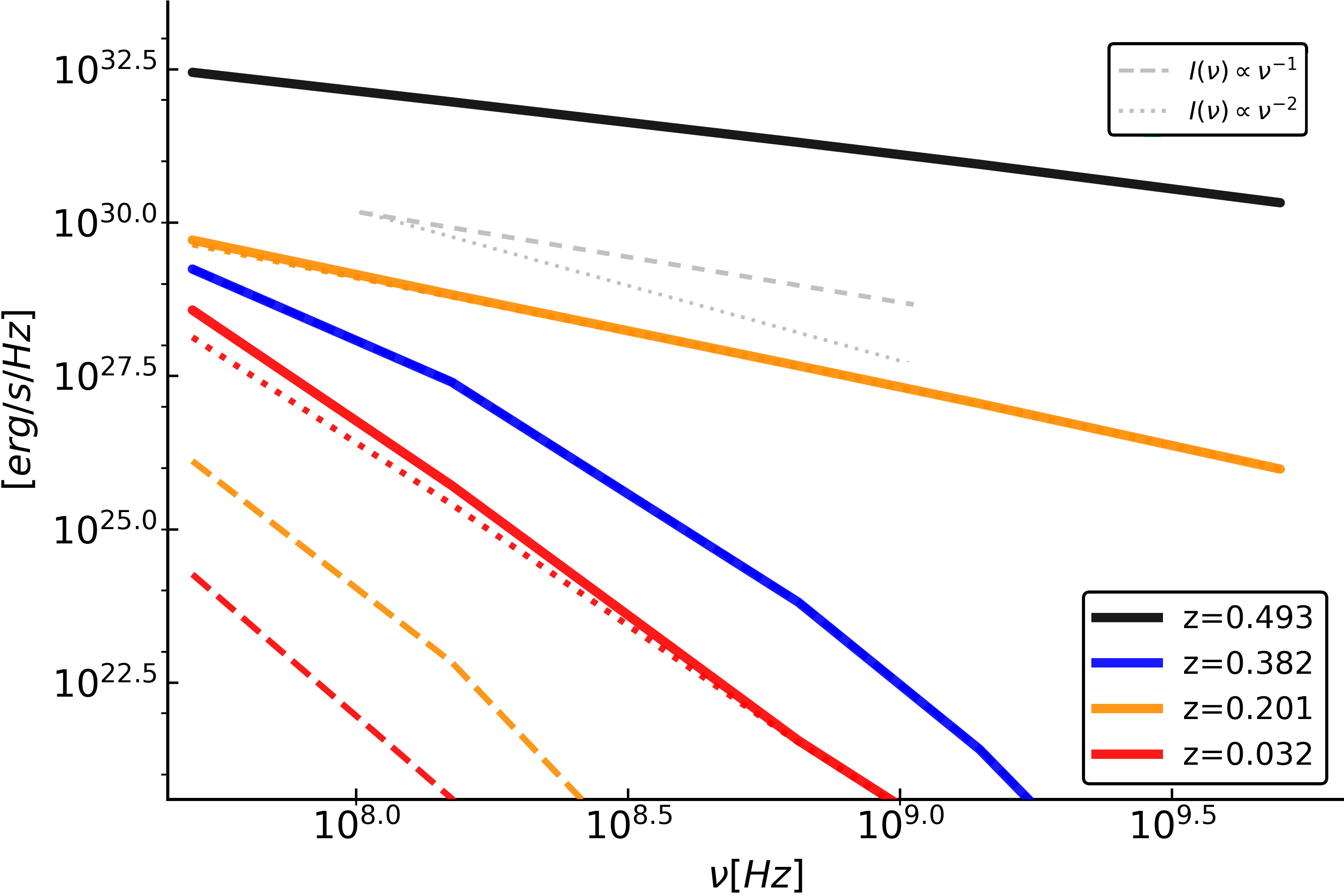} 
\end{center}
\caption{Left panel: electron momentum spectra for four different epochs and for all tracers located within a $r \leq 1 ~\rm Mpc$ radius from the moving cluster centre, considering the evolution of re-accelerated electrons seeded by radio jets, under the three investigated model for CRe energy evolution (C, CS and CST).  Right panel: evolving radio emission spectra for the same families of electrons and epochs (same meaning of the different line styles - the three models give the same radio spectrum for the first two epochs so all model lines are super imposed there), without observational cuts. The thin grey  lines give two different radio slopes ($I(\nu) \propto \nu^{-1}$ and $\propto \nu^{-2}$) to guide the eye.}
\label{fig:spec1}
\end{figure*}

\begin{figure*}
\begin{center}
\includegraphics[width=0.6\textwidth]{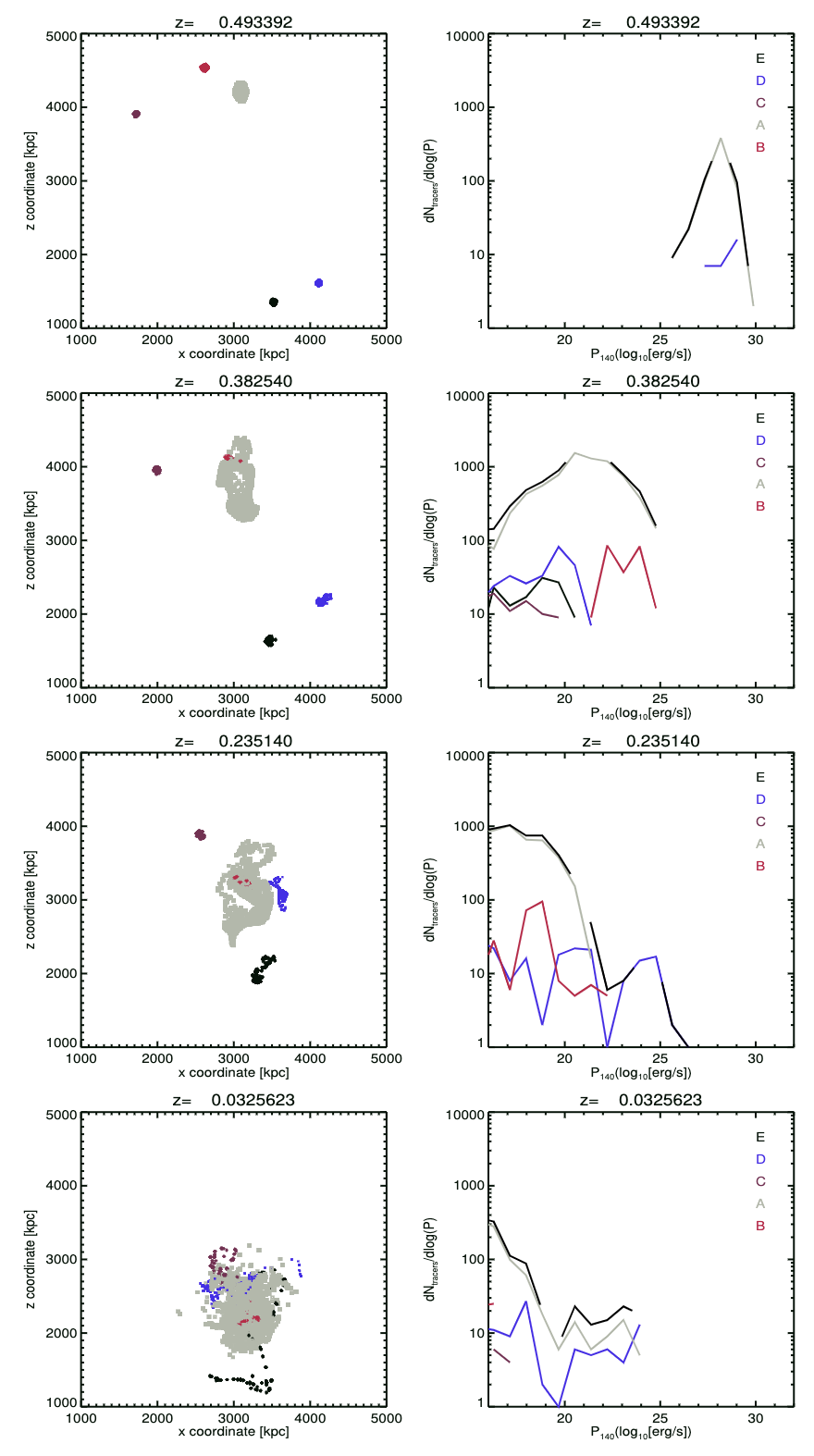}
\end{center}
\caption{Left panels: evolving projected location of simulated populations of relativistic electrons seeded by radio jets. Right panels: distribution of radio emission power at 140 MHz for the corresponding families of tracers (in the CST model), and for their total (grey dashed lines).}
\label{fig:families}
\end{figure*}  
\section{Results}
\label{results}

\subsection{The evolution of electrons seeded by radio jets}
\label{radio_jets}
We begin by analysing the evolution of the three-dimensional location of the electrons seeded by radio jets. The time sequence shown in Fig.~\ref{fig:movie1} shows the evolving gas density (green colors), mass weighted gas temperature (blue) and the 
the projected location of all electrons injected by radio jets (red) starting from $z=0.5$ and for a few interesting evolutionary steps, down to $z=0.032$. 

Electrons rapidly expand into the cluster atmosphere, and get progressively dispersed and mixed on larger scales, by being entrained by turbulent motions or large-scale advection flows induced by the accretion of cluster satellites. 

With the exception of the central, dominant SMBH (which corresponds to the BCG in this system), the initial jet power from the other galaxies is not sufficient to push electrons to large distances from their host SMBH, and electrons first settle into bubble-like distributions, to be later dispersed into the main cluster while they follow the accretion of their host galaxy. 
The relative motions between the injected electrons and the ICM are large enough to thoroughly disperse the CRe throughout the cluster volume. By the end of our simulation, a large fraction of the cluster area (i.e. the cluster volume seen in projection) appears to be filled by the electrons seeded by a single burst of our radio galaxies (we will quantify the volume filling factor of these electrons in the next Section). 

\begin{figure*}
\begin{center}
\includegraphics[width=0.8\textwidth]{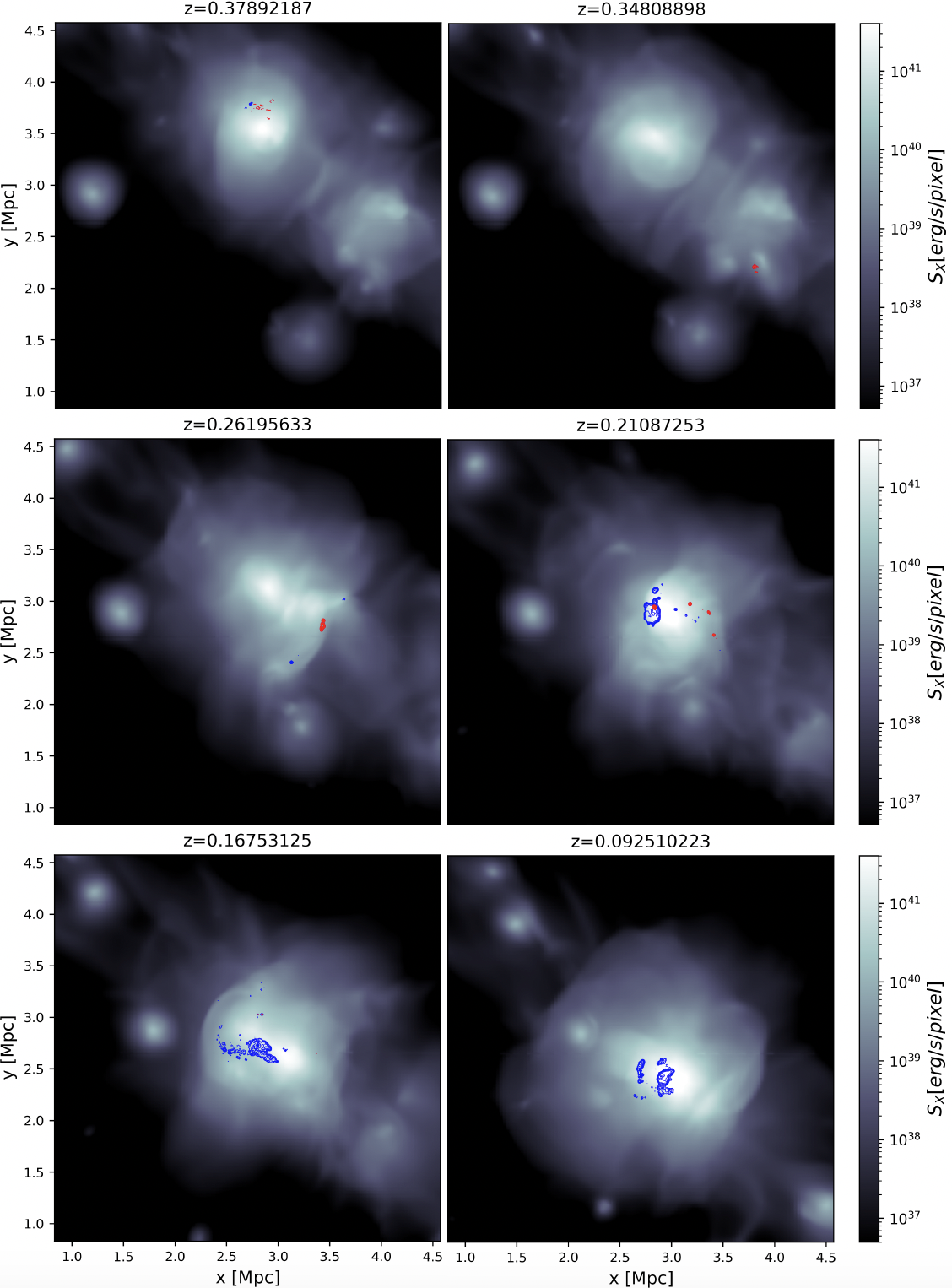}

\end{center}
\caption{Evolution of the X-ray surface brightness (grey colors) and of the detectable radio emission at 50 MHz with LOFAR LBA for $z \leq 0.4$, for all electrons seeded by radio jets at $z=0.5$ and later re-accelerated by shocks and turbulence (CST model, red contours), or also considering the additional contribution from electrons seeded by merger shock waves  from $z=0.4$ (CST model, blue contours). The full movie of this sequence can be watched at https://youtu.be/MPEFiIzr-V8.} 
\label{fig:movie2b}
\end{figure*}

The fuelling of electrons in the innermost cluster region is thus largely dominated, at least for the first $\sim 0.8~\rm Gyr$ ($z\sim 0.5-0.4$) by the contribution of the central radio galaxy only. 
Fig.~\ref{fig:movie2} shows the projected radio emission at 50 MHz in the first stage of radio lobe expansion, i.e. from $z \approx 0.5$ to $z \approx 0.4$, shortly after which the remnant emission even from the most powerful radio galaxy becomes invisible (the rest of the evolution of the radio detectable emission is shown in Fig.\ref{fig:movie2b}, in which we also add the emission from shock injected electrons). 
To estimate which emission would be detectable, we considered a fixed luminosity distance of $d_L=132 \rm ~Mpc$ for all snapshots, in which case our simulated pixel size corresponds to the resolution beam of LOFAR Low Band Antenna (LBA) at 50 MHz, of $\approx 8 \rm ~kpc$, considering a beam of $\theta=12.5"$, and a sensitivity of $\sigma = 5.7 \cdot 10^{-4} \rm ~Jy/beam$, as in \citep[][]{2022SciA....8.7623B}. 

After the initial bright stage of jet emission, 
only a tiny fraction of the emission from the injected electrons remains visible at radio frequencies (red contours). Already after 0.5 Gyr since the injection (z=0.436 in the images), the only visible emission is from the stretched remnant lobe structures, which have begun to mix in the innermost cluster atmosphere. This mixing occurs mostly in the vertical direction in the innermost cluster atmosphere, thus carrying memory of the initial jet orientation. The detectable structure are mostly filamentary and patchy, with $\leq 500 ~\rm kpc$ length, until $z=0.38$, after which there is no detectable radio emission for a long while.
However, the progressive mixing and dispersal of the different families of electrons, now become a "fossil" population, continues, and later on (e.g. $z \approx 0.37$, $0.26$, $0.21$ and in Fig.\ref{fig:movie2b}) patches of electrons that are crossed by shock waves, or are intersected by turbulence, get re-accelerated and become visible again. All these events are triggered by accretion events of substructures, which mostly happens along a diagonal in the image, and often release  pairs of weak ($\mathcal{M} \leq 3$) shock waves, and stir new subsonic turbulence in the main cluster centre. 

The electron momentum spectra and the radio spectra (of all tracers located within a $r \leq 1 ~\rm Mpc$ radius from the moving cluster centre) at different epochs, shown in Fig.~\ref{fig:spec1},
show that only when Fermi I and Fermi II terms are included (models CS and CST) the electron spectra remain energetic enough to produce significant radio emission, typically with very steep ($\alpha \leq 1.5-2$) radio spectra. Only at late times, turbulent re-acceleration is important to increase the budget of low-energy electrons and their low-frequency emission (CST). The net effect of turbulent re-acceleration is to systematically increase the budget of low energy electrons ($p \leq 10^3$), as well as the low-frequency radio emission ($\nu \leq 610$ MHz), beyond the effect on shock re-acceleration alone, as showed by the small excess of the solid lines compared to the dotted lines (otherwise always superimposed) in Fig.~\ref{fig:spec1} . However, it is difficult to state in a quantitative way the relative effects of turbulent re-acceleration in comparison with shock re-acceleration: we studied in detail in \citet{va21b} that this depends on the epoch of observation, and on the specific dynamical history experienced by tracers, with a range of extra energy below $p \leq 10^3$ which can range from $\sim 2$ to $\sim 10$ depending on the time since the last shock event experienced by tracers. 

We notice that, especially in the initial stage of lobe expansions, where the further injection of CRe from shocks is small, the fraction of the CRe population in the radio band may slightly change for a higher assumed initial injection fraction of CRe electrons in jets. However, our choice ($0.5\%$ of the number density of thermal particles) already is at the high end of the range typically assumed for these sources \citep[e.g.][]{oj10,2012ApJ...750..166M}, and moreover a $\sim 10$ higher normalisation of the initial CRe density would make our central BCG extremely powerful, e.g. nearly at the top of the distribution of radio power recently measured by \citet[][]{pasini21}, which would be odd for a source in such a small mass cluster of galaxies.

Although the electrons injected by the central BCG clearly dominates the radio emission at the start of the simulation, its relative contribution when re-acceleration processes become dominant changes over time. 
Figure~\ref{fig:families} shows the trajectories of particles initially belonging to different radio galaxies (with different colors) and their relative contribution to the total radio emission from the cluster (here without observational cuts for the CST model). 
While the emission is initially dominated by the powerful "A" source, the relative contribution of the electrons in its lobe mixes with time with the electrons coming from other radio sources.  At later times also the pool of electrons released by source "D" and "B"  become the bright emitters in the field since their particles get crossed by shock waves. This shows that the seeding of electrons from sources other than the central powerful radio galaxy likely have a key role in enriching the ICM with additional fossil electrons, and it thus complements our previous results, which were limited to the impact of a single radio source in the cluster \citep[][]{va21b,va22}. 
Moreover, detectable radio emission features can arise from the blending of the initially distinct population of radio electrons, similar to what has been proposed to explain a few puzzling steep spectrum radio sources \citep[][]{Hodgson21a}.
However, the size of the detectable regions are very small, i.e. $\leq 100$ kpc at all epochs for $z \leq 0.4$ (see Fig.\ref{fig:movie2b}). This means that in all our tests so far, it appears not plausible that the much larger extent of giant radio relics ($\sim 2 \rm ~Mpc$ in the largest cases, e.g. \cite{vw10,rajp20}) can be efficiently filled by a uniform pool of fossil electrons, and thus provide alone the background of seed electrons for Fermi I re-acceleration in these objects. On the other hand, the seeding of electrons from radio galaxies appear a more plausible channel to refill the central regions of classic radio halos, as already partially supported by direct observations\citep[e.g.][]{2018MNRAS.473.3536W,2020ApJ...897...93B}. 
Of course, these result has to be considered only as tentative so far, given the single jet events we could simulate so far.

\begin{figure*}
\begin{center}

\includegraphics[width=0.4\textwidth]{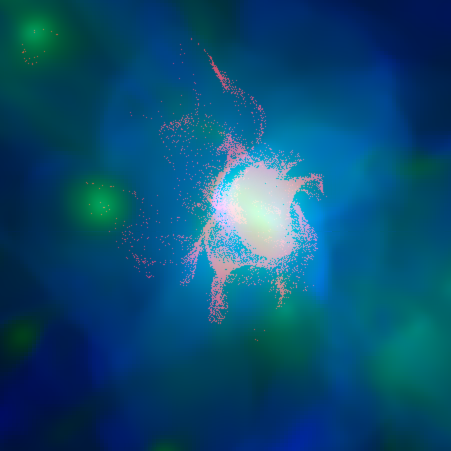}
\includegraphics[width=0.4\textwidth]{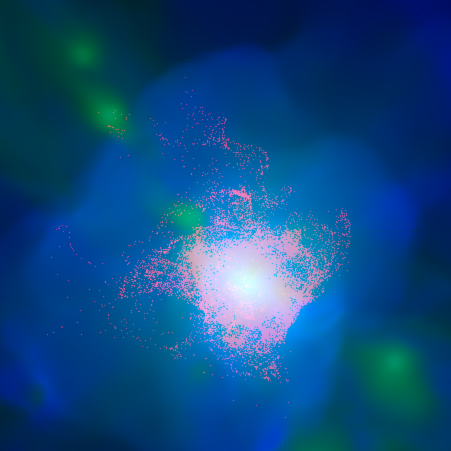}
\end{center}
\caption{Red-green-blue composite images showing two evolved snapshots (z=$0.102$ and $0.032$, respectively) of the model in which electrons are seeded by merger shocks, starting from $z=0.4$.  Each image is $5 \times 5$ Mpc$^2$ across and it shows the mass-weighted gas temperature along the line of sight (blue), the projected gas density (green) and the location of electron tracer, regardless of their energy (red). These two panels match the final two snapshots of the ones in Fig.~2.}
\label{fig:movie1b}
\end{figure*}

\begin{figure*}
\begin{center}
    \includegraphics[width=0.92\textwidth]{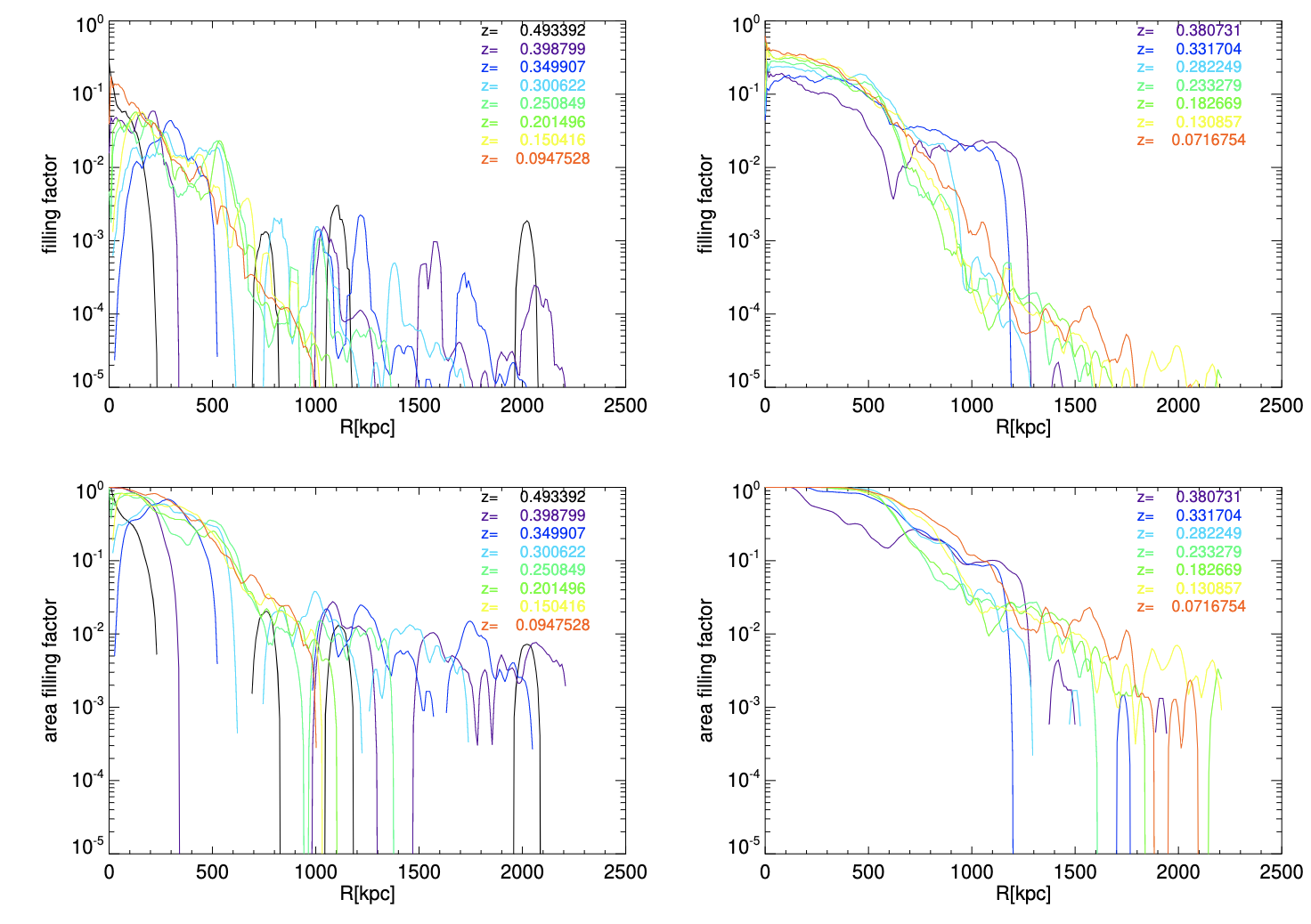}
\end{center}
\caption{Evolution of the volume filling factor (top row) or of the projected area filling factor (bottom),  for either 3-dimensional or 2-dimensional radial shells centred on the (moving) gas density peak of our simulated cluster of galaxies, for the electrons seeded by radio jets at $z=0.5$ (left panels) and for the electrons seeded by merger shocks from $z=0.4$ (right panels).}
\label{fig:filling}
\end{figure*}  

\subsection{Shock injection of electrons versus the injection by radio jets}
\label{shocks}
Next, we compare with the result for electrons seeded by merger shocks, in which we track the additional contribution from electrons injected by merger shock waves starting from $z=0.4$, to have a more conservative view of what shocks driven by accretion alone (i.e. without considering shocks triggered by AGN events) can do. 

Even if this includes more CR electrons in our modelling, it still represent an underestimate, because we only track here the contribution from merger shocks in the innermost cluster regions, and neglect the injection of electrons from shocks in more peripheral regions. 
Indeed, given the large number of tracers to follow with our electron solver, we initialised $\approx 2 \cdot 10^5$ particles limited to $T \geq 10^7 \rm K$ cells (approximately within $\leq R_{\rm 500}$ for this system at that epoch), which is the environment in which merger shock develops and that is also expected to dominate the injection of CR in clusters \citep[e.g.][]{scienzo16}. 
Therefore, if even more peripheral shocks would be accounted for in the seeding of electrons, the dominance of shock-seeded electrons over radio jets-seeded electrons would only increase. On the other hand, it is still possible that the time integrated effects of seeding by radio jets, considering their entire duty cycle even from high redshift, can produce volume filling factors approaching the ones by electrons seeded by merger shocks. 

The volume spanned by merger shocks is a significant fraction of the cluster volume, and hence already after $\sim 1$ Gyr since this mechanism is allowed in our simulation, a large portion of the innermost ICM is filled with some amount of cosmic ray electrons, as shown by Fig.\ref{fig:movie1b} which displays a much larger area covering factor of electrons, compared to the corresponding maps for the seeding from radio jets (last two panels of Fig.\ref{fig:movie1}). 

We computed the  3-dimensional and 2-dimensional filling factors  for our electron tracers released from the different mechanisms as a function of time.  Figure \ref{fig:filling} shows the evolution of the fraction of volume, or projected area, of radial shells (centred in the moving cluster density peak) filled by at least one simulated tracer.
The two trends are opposite and clear.
Electrons seeded by radio jets (and in particular by the central one, corresponding to the BCG) occupy a large fraction of the central cluster volume right after their injection epoch, and get advected to a larger volume, and their volume filling factor drops to  $\sim 0.1-1\%$ in the innermost $1 \rm ~Mpc^3$ region, at most epochs.
Conversely, electrons seeded by shocks are initially deposited preferentially in outer radial shells in the cluster, where shocks are formed, but then  steadily increase their volume filling factor over time, reaching a $\sim 3-30\%$ filling in the innermost $1 \rm ~Mpc^3$ already at $z \sim 0.23$, i.e. $\sim 1.5$ Gyr since their are first injected in our model. 
More relevant for the potential radio observations which can target such low energy electrons, the filling factors for the projected surface are higher in both cases, but again the filling factor of electrons seeded by shocks is much higher at later times, i.e. $\geq 5-10$ more projected surface is filled with fossil electrons seeded by shocks at all radii, for $z \leq 0.23$.

Again, the latter number is probably underestimated by neglecting the entire injection process of CR electrons by structure formation shocks \citep[e.g.][]{pi10}.

Since seed cosmic-ray electrons are spread over a larger volume, they also have more frequent chances of being re-accelerated by subsequent shocks sweeping the same system, and produce radio detectable emission, as shown by the sequence of radio images in Fig. \ref{fig:movie2b}, in which we applied the same observational cuts of Fig.\ref{fig:movie2}. While fossil electrons released early on by radio jets are sometime re-accelerated by shocks and turbulence at late redshift, the detectable emission covers a wider area if electrons injected by shocks are included. In the latter case, they more naturally give rise to a few $\sim 500$ kpc large relics in our simulation, with a more uniform surface brightness and radio spectra. Considering that here we can only track the evolution of particles first injected at $z=0.4$ in the innermost cluster regions, it is quite plausible to expect that the inclusion of all past injection events of electrons should produce even larger scales, and possibly smoother, radio emission regions. 

\begin{figure*}
\begin{center}

     \includegraphics[width=0.45\textwidth]{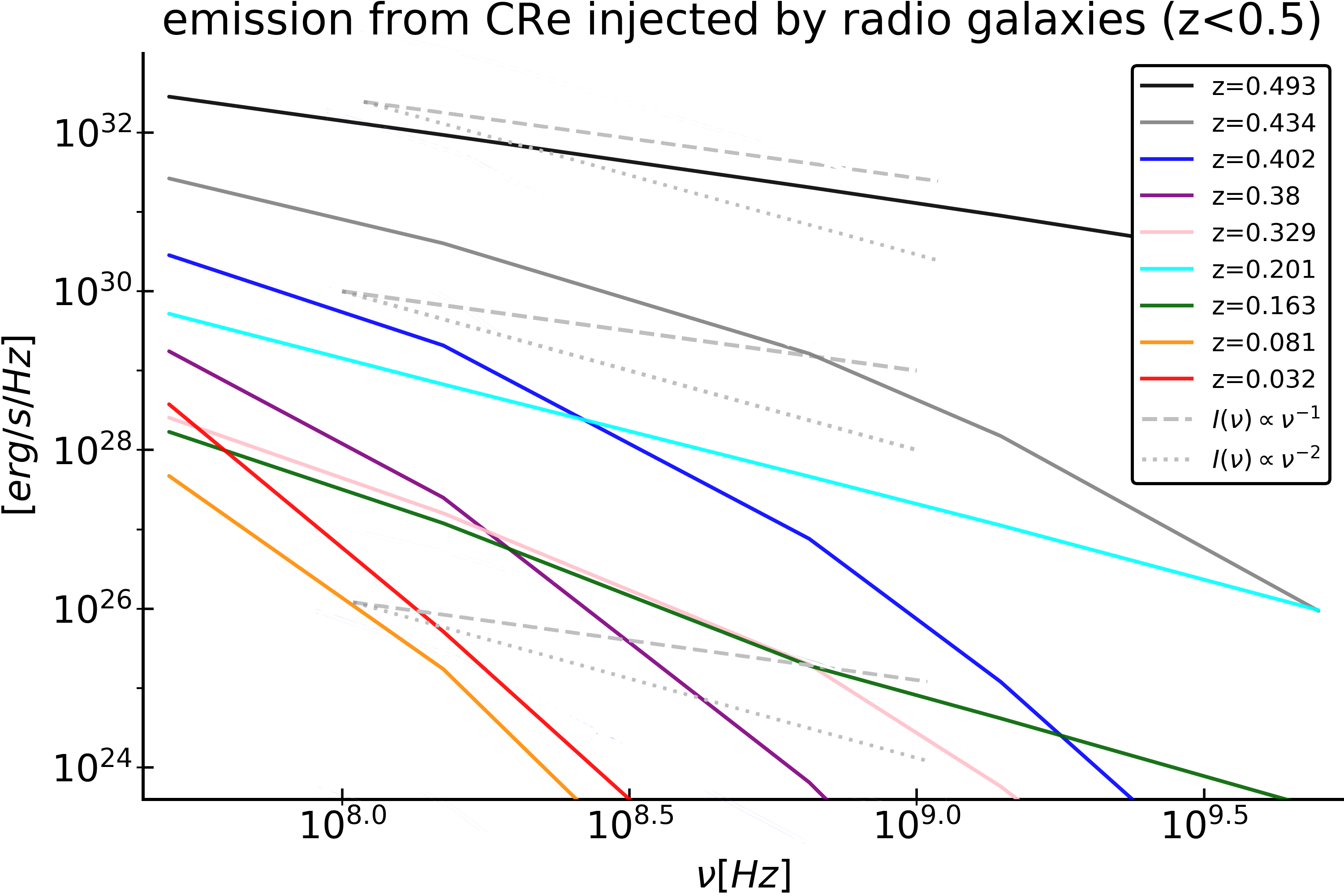}
     \includegraphics[width=0.45\textwidth]{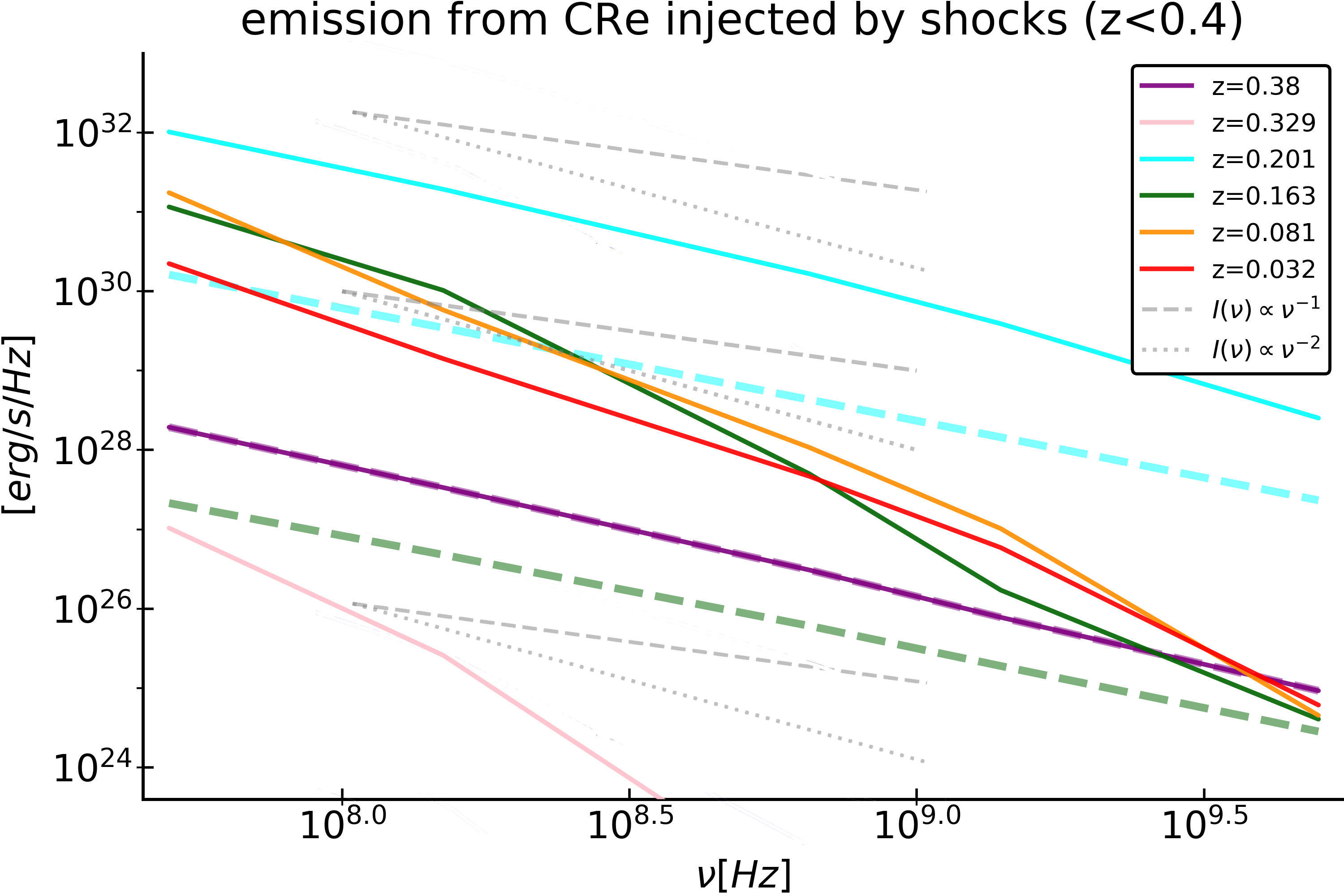}
\end{center}
%\caption{Left panel: electron momentum spectra for four  different epochs, considering the evolution of re-accelerated electrons  injected by internal merger shocks starting from $z=0.4$. The dashed lines show the power-law distribution of electrons freshly injected by shocks at the given epoch. Right panel: 
\caption{Comparison of the evolving radio emission spectra from our simulation form the two investigated scenarios for the origin of cosmic ray electrons (without considering observational cuts). Left panel: emission from electrons seeded by radio jets at z=0.5 (CST model). Right panel: emission from electrons seeded by merger shocks, starting from $z=0.4$ (CST model). The additional thick dashed lines show the radio emission only coming from electrons freshly injected by shocks at the epoch of observation (these lines are only visible for the snapshots with enough ongoing shock acceleration. The repeated thin grey dotted lines give two different radio slopes ($I(\nu) \propto \nu^{-1}$ and $\propto \nu^{-2}$) to guide the eye.}
\label{fig:spec2}
\end{figure*}

A final comparison of the long-term radio evolution of emission from electrons seeded by merger shocks, and the one by radio jets is given  in Fig. \ref{fig:spec2}. 
Following from the previous trends of filling factors, while the contribution from electrons seeded by radio jets overall decreases with time, even in the presence of re-acceleration events, the one from electrons seeded by shocks overall increases with time, and by the latest epochs in the simulation the total radio emission from the cluster is $\sim 10^2$ larger at $50$ MHz if electrons injected by shocks are considered. Given the higher frequency of re-acceleration events experienced by the volume filling population of shock-injected electrons, the emission from the latter is at most epochs significantly flatter than the one coming from re-accelerated electrons from radio galaxies. 

Finally, the right panel also shows the radio emission only for the prompt injection of electrons by shocks (thick dashed lines). While at the beginning of the sequence the contribution from freshly injected electrons dominates the total budget, this becomes over time secondary with respect to the effect of shock re-acceleration. Limited to this particular system, and its history of weak internal shocks ($\mathcal{M} \leq 3$), this is in line with the frequent finding that radio relic emission observed in many clusters of galaxies is better explained by some form of re-acceleration of fossil electrons \citep[e.g.][]{pi10,ka12,2019MNRAS.489.3905S,rajp21a}, which in this case appear to be most likely a multiple shock scenario \citep[e.g.][]{2022MNRAS.509.1160I}. On the contrary, observed radio relic emission compatible with the "simple" prompt seeding of electrons by shocks, based on DSA, appears to be compatible only with a very few objects observed in the radio band \citep[e.g.][]{2020MNRAS.496L..48L}.

%%%%%%%%%%%%%%%%%%%%%%%%%%%%%%%%%%%%%%%%%%
%\section{Discussion} \label{sec:discussion}
%%%%%%%%%%%%%%%%%%%%%%%%%%%%%%%%%%%%%%%%%%
\section{Discussion \& Conclusions}
\label{sec:conclusions}

In this work, which is a follow-up study after our recent \citet{va21b} and \citet{va22} work, we presented a first assessment of the relative role of multiple radio jets, and of merger shocks waves, in  a) filling the  ICM with fossil relativistic electrons, and b) giving rise to detectable radio emission, based on state-of-the-art models of particle (re)acceleration. 

%Compared to previous work, we simulated here the simultaneous activity by eight radio galaxies producing a feedback event at the same time ($z=0.5$) within the same group of galaxies.
%This is of course an extreme case, considering that it is rare to find such a large number of object active at the same time in a group of galaxies (CONFIRM), yet recent low frequency surveys have established that the most massive radio galaxies in clusters and groups of galaxies are "on" most of their lifetime (Sabater+). For this reason, our model can be considered the extreme power event of a generally continuous activity of radio galaxies in these environment. 

%Under these extreme conditions, we focused on two main questions, which are of great interest for the study of diffuse radio sources (e.g. radio halos, radio relics and mega halos) in galaxy clusters, namely: a) which fraction of the cluster volume is enriched with "fossil" electrons injected by radio lobes, as a function of time and b) how does the presence of fossil electrons affect the later emission from radio relics?

Based on the simulation of a single system, our most significant conclusions can be summarised as follows: 

\begin{itemize}
    \item The seeding of electrons from sources other than the central powerful radio galaxy have a significant role in enriching the ICM with an additional amount of fossil electrons. During the evolution, powerful enough radio emission features can be produced from populations of electrons seeded by peripheral galaxies, which also blend and mix over time.
    
    \item Right after the active stage of radio jets, the remnant radio plasma typically dominates the radio emission at a $\leq 0.5$ Mpc distance from sources, even up to $\sim 0.5-1$ Gyr since its first injection, but only if quasi-continuous re-acceleration events (e.g. frequent weak shocks or turbulence) are active on particles. The emission from re-accelerated electrons seeded by radio galaxies can produce detectable emission at any time, in presence of re-acceleration events, but only leading to small  ($\leq 100$ kpc in our test) and often filamentary radio features with steep radio spectra ($\alpha \geq 1.5-2$). This makes it  hard for radio galaxies alone to fuel giant radio relics on scales of $\sim 1-2 $ Mpc, with the required uniform population of fossil electrons, while they appear a sufficiently viable channel to fuel the regions of central radio halos. 
    
    \item If merger shocks (here limited to those forming in $T \geq 10^7 \rm K$ gas, for computational limitations) also contribute seeds of cosmic rays, then the volume filling factor of fossil electrons in the central $r \leq 1 \rm ~Mpc$ region increases from $0.1-1\%$ (in the case of CR electrons only injected by radio galaxies) to $3-30\%$ if also shock-injected CR electrons are included.  The injection by shocks occurs on much larger scales than injection by radio galaxies. It can also naturally lead to large and correlated populations of fossil electrons that subsequent re-acceleration events can illuminate on $\sim \rm Mpc$ scales and with flatter spectra than the one from radio galaxies, owing to the shorter time elapsed between re-acceleration events.
    \end{itemize}

A number of limitations prevent us from generalising our results, that are a first attempt at modelling diffuse radio emission in clusters, caused by the seeding by radio galaxies.

First, our prescription for radio galaxies is simplistic, in the sense that we only considered relatively low-power jests, activated all at the same time and only once in the late cluster evolution. Radio galaxies with higher powers can inject more fossil electrons, even if our tests in \cite{va21b} and \citet{va22} actually showed that too powerful jets tend to spread electrons too widely. On the other hand,  recent low-frequency surveys have established that the most massive radio galaxies in clusters and groups of galaxies are "on" most of their lifetime \citep[][]{2019A&A...622A..17S}.  
We also did not include self-regulated feedback and the effect of multiple AGN bursts, leading to repeated radio jet activity, as observed \citep[e.g.][]{brienza22}.
While it is plausible that the collective effect of all above missing ingredients can significantly increase the volume filling factor of fossil electrons from radio galaxies, it still
remains to be proven that radio galaxies alone can fill with seed CR electrons the  $\sim \rm 1-2~ Mpc$ wide radio relics, and result into the uniform spectral and brightness distribution observed in the most spectacular examples. Our results implies already that 
the single activity of just a few radio sources is unlikely to do that. 

Second, discretisation effects prevent us from following the additional diffusion of electrons on small scales. However, this should not dramatically alter the mixing pattern captured by our tracer approach. But it will increase the filling factor that we can measure (hence lower limit). Our tracer technique is suitable to follow most of the turbulent diffusion acting on cosmic rays, considering that the propagation of tracers in sensitive to the largest eddies ($\sim 100-200 \rm ~kpc$) resolved in the turbulent cascade by the adopted MHD solver \cite[e.g.][]{va10tracers,wi17b}. However, our numerical method cannot capture turbulent diffusion on the smallest scales, which can happen on timescales of the same order of our cooling and re-acceleration timescales. The CR diffusion timescale can be roughly quantified in the hydro regime as $l_{\rm D} \sim 2 \sqrt{D_{\rm turb,small} \cdot t}$, where the small-scale turbulent diffusion coefficient is $D_{\rm turb,small} \sim (V_{\rm turb,small} \cdot \Delta x)/3$ (with $\Delta x$ and $V_{\rm turb,small}$ being, respectively, the effective numerical resolution at the tracer location and the turbulent velocity fluctuation within that scale). Based on our previous work, in which the same cluster was simulated, we can estimate a typical rms turbulent velocity of $V_{\rm turb,small} \leq 5 \rm ~km/s$ for  $\Delta x \leq 8-16 ~\rm kpc$, which gives $l_{\rm D} \leq 0.01 \rm ~kpc$ over our typical timestep $\delta t \approx 32 \rm~ Myr$, or $\sim 2 \rm ~kpc$ for the entire simulated evolution. 
However, despite the fact that the effect of unresolved CR diffusion should be small and subdominant compared to that of turbulent diffusion, a conclusive view of the circulation of CR electrons in the simulated ICM requires a fully fluid treatment of CR  \citep[e.g][]{nolting19a}.
Moreover,  a few other obvious injectors of fossil radio electrons are missing in our model, most noticeably the injection from star formation winds \citep[e.g.][]{1991A&A...245...79B,2018ApJ...856..112F,2022arXiv221014232B} whose impact is observed at radio frequencies \citep[e.g.][]{2021Ap&SS.366..117H}.

Finally, we must remark that the applicability of our conclusions is limited to the fact that we only analysed one cluster of galaxies, which prevent us to monitor how the transport and ageing  of particles change as a function of the host cluster mass and, probably most importantly, on the host cluster dynamical state. The latter is important, because of the striking observed association between disturbed and merging clusters of galaxies and central radio halos or radio relics, and between more relaxed and cool-core clusters of galaxies with mini-halo emission \citep[e.g.][]{2019SSRv..215...16V}. Our previous works, employing Lagragian tracers, have showed that the turbulent mixing and dispersion of tracers tracking mass accretions of simulated clusters with different dynamical histories are nearly indistinguishable after a few  $10^2 \rm ~Myr$ \citep[][]{va10tracers}, and so is the typical statistics of turbulence and shocks recorded along their propagation history \citep[][]{wi17b}. However, we did found some significant differences in the mixing pattern of tracers initially located within the core of already formed relaxed or perturbed clusters of galaxies, and meant to track the mixing of metals with the ICM. Our analysis indeed measured a smaller degree of mixing of tracers outside of the cores of the most relaxed clusters in the simulated sample, even after a few Gyr of evolution \citep[e.g.][]{va10tracers}. However, no AGN feedback was modelled in those early works, and it is likely that this artificially reduced the amount of turbulent and mixing motions in the core of the most relaxed clusters. Only with future resimulations we will be in the position of generalizing the trends suggested by the work presented here.

In summary, our results show that in order to properly model the long-term evolution of radio emission in clusters and groups it appears necessary that all key ingredients of cosmic ray enrichment are included. Their relative importance can vary with spatial scale, proximity to sources, distance from the cluster centre and redshift.
This, in turn, poses new challenges to the numerical modelling, as well as to the increased computational costs to co-evolve the thermal and non-thermal components of the ICM - jointly with the pervasive magnetic fields which link the two.

%%%%%%%%%%%%%%%%%%%%%%%%%%%%%%%%%%%%%%%%%%
\vspace{6pt}

%%%%%%%%%%%%%%%%%%%%%%%%%%%%%%%%%%%%%%%%%%
\authorcontributions{F.V. prepared the manuscript and all the numerical simulations analysed in this article. D.W. produced the lagrangian history of tracer particle used in this work. G. B. and  M.B. contributed to the design and testing of the relativistic electron solver used for this work  All authors  contributed in the drafting, editing and literature review of the manuscript and in the scientific interpretation of results.}

%%%%%%%%%%%%%%%%%%%%%%%%%%%%%%%%%%%%%%%%%%

\conflictsofinterest{The authors declare no conflict of interest.}

\funding{F.V. acknowledges financial support from the Horizon 2020 program under the ERC Starting Grant {\magcow}, no. 714196. 
D.W. is funded by the Deutsche Forschungsgemeinschaft (DFG, German Research Foundation) - 441694982.
G.B. acknowledges partial support from mainstream PRIN INAF "Galaxy cluster science with LOFAR".   MB acknowledges support from the Deutsche Forschungsgemeinschaft under Germany's Excellence Strategy - EXC 2121 “Quantum Universe” - 390833306. }

\institutionalreview{Not applicable.}

\informedconsent{Not applicable.}

\dataavailability{In this work we used the {\enzo} code (\hyperlink{http://enzo-project.org}{http://enzo-project.org}), the product of a collaborative effort of scientists at many universities and national laboratories. 
We share the public version of our solver for relativistic electrons at the URL \url{https://github.com/FrancoVazza/JULIA/tree/master/ROGER}.}

\acknowledgments{F.V. gratefully acknowledges : a) the Gauss Centre for Supercomputing e.V. (www.gauss-centre.eu) for supporting this project by providing computing time through the John von Neumann Institute for Computing (NIC) on the GCS Supercomputer JUWELS at J\"ulich Supercomputing Centre (JSC), under projects "radgalicm2" and b)  the Swiss National Supercomputing Centre (CSCS), Switzerland.", for the access to the Piz Daint, under the allotted project "s1096".  \\ }

%%%%%%%%%%%%%%%%%%%%%%%%%%%%%%%%%%%%%%%%%%

\abbreviations{The following abbreviations are used in this manuscript:\\
\noindent
\begin{tabular}{@{}ll}
AGN & Active Galactic Nucleus\\
AMR& Adaptive Mesh Refinement\\
ASA& Adiabatic Stochastic Acceleration\\
ASKAP& Australian Square Kilometre Array Pathfinder\\
BCG& Brightest Central Galaxy\\
CR& Cosmic Rays\\
DM& Dark Matter\\
DSA&Diffusive Shock Acceleration\\
%CMB& Cosmic Microwave Background\\
%FRB& Fast Radio Bursts\\
%GLEAM& The Galactic and Extra Galactic All Sky MWA Survey\\
%GPU& Graphics Processing Unit\\
HBA& High Band Antenna\\
HLL& Harten-Lax van Leer\\
ICM& Intra Cluster Medium\\
%IGM& Inter Galactic Medium\\
%IR& Infa Red\\
%JVLA& Karl G. Jansky Very Large Array\\
%KAT& Karoo Array Telescope\\
LBA& Low Band Antenna\\
LOFAR& Low Frequency Array\\
%LOS& line-of-sight\\
MAGOCW& The Magnetised Cosmic Web\\
%MeerKAT& MeerKAT \\
MHD& Magneto Hydro Dynamics\\
MWA& Murchison Widefield Array\\
PLM& Piecewise Linear Method\\
RK& Runge-Kutta\\
%RM& Rotation Measure\\
%SDSS& Sloan Digital Sky Survey\\
%SKA& Square Kilometer Array\\
SMBH& Super Massive Black Hole\\
SPH& Smoothed Particle Hydrodynamics\\
%TVD& Total Variation Dininishing\\
%UHECR& Ultra High Energy Cosmic Rays\\
%WHIM& Warm Hot Interagalactic Medium\\
$\Lambda$CDM& Lambda Cold Dark Matter
\end{tabular}}

%%%%%%%%%%%%%%%%%%%%%%%%%%%%%%%%%%%%%%%%%%

\reftitle{References}
\externalbibliography{yes}
\bibliography{franco3,dan}
\end{paracol}
\end{document}